\documentclass{article}
\usepackage{graphicx}
\begin{document}
\title{{\bf A new kind of science}}
\author{Alex Hansen\footnote{PoreLab, Department of Physics, Norwegian University of Science and Technology, N--7491 Trondheim, Norway.} and Sauro Succi\footnote{Italian Institute of Technology, 
Viale Regina Elena, 291, I--00161, Rome, Italy, Physics Department, 
Harvard University, Cambridge USA and PoreLab, Department of Physics, Norwegian University of Science and Technology, N--7491 Trondheim, Norway.}}

\maketitle
%----------------------------------------------------
\begin{abstract}
We discuss whether science is in the process of being transformed from a 
quest for causality to a quest for correlation in light of the recent 
development in artificial intelligence.    
We observe that while a blind trust in the most seductive 
promises of AI is surely to be avoided, a judicious combination of computer
simulation based on physical insight and the machine learning ability
to explore ultra-dimensional spaces, holds potential for 
transformative progress in the way science is going to
be pursued in the years to come.
\end{abstract}
%----------------------------------------------------
\section{Introduction}
\label{intro}

On October 8, 2024 the Nobel Prize in Physics was awarded to John J.\ Hopfield
and Geoffrey E.\ Hinton with the following motivation 
``for foundational discoveries and inventions that enable machine 
learning with artificial neural networks”
\cite{nobel-phys}.

The day after the Chemistry Nobel was awarded to David Baker 
(Washington University) for ``for computational protein design," and
to Demis Hassabis and John M.\ Jumper of DeepMind for ``for protein structure prediction" \cite{nobel-chem}.

John Hopfield was saluted with great satisfaction by 
the (statistical) physics community as a champion of that kind 
of interdisciplinary statistical physics that decisively impacts on different 
fields, in this case biology and neuroscience.
Geoffrey Hinton came as a bit of a surprise, as he is a highly 
distinguished authority but rather in computer science 
than physics, as witnessed by his recent award of the
Turing Prize, along with Yan LeCun and Joshua Bengio, for 
their groundbreaking work on neural networks.
Yet, Hinton's work certainly related to physics, especially 
his celebrated recursive Boltzmann machines.

The Nobel Prize in Chemistry, on the other hand, came as a direct 
hommage to Artificial Intelligence (AI), a statement 
which is particularly true for the DeepMind winners.

In short, the main critique is that, at variance with most bombastic headlines, 
AlfaFold is a monumental and extremely impressive {\it tour de force\/} 
of computer science and engineering, but does not really "solve" the
protein folding problem.
  
many scientists think it does not, mainly because it did not deliver any 
real insight into the phenomenon of protein folding, namely
the dynamics taking from primary structures to their native form.
The point may seem far-fetched and kind of artificial, but it is not. 
Protein folding is a dynamical process and if we are to gain useful insights to
cure neurological diseases, we need not just the end points but the entire
trajectory, i.e., the dynamics. Maybe in five or ten years from now AlfaFold
will bridge this gap, but till then the claim that the protein folding problem
is cracked, is simply overstated.

The reactions are hot and split: for some this is the end of Galilean science,
as provocatively announced in C.\ Anderson's 2008 Wire Magazine article 
\cite{a08}, namely the triumph of Correlation over Causation (for quick and
direct counter-arguments see \cite{CH,SC19}). 
For others, it is a mere fact of life that Machine Learning (ML) algorithms, no matter 
how  empirical, manage to capture levels of complexity unattainable 
by any other method, including our most powerful theories and computer simulations.
One may find a weak echo of this debate in that which surrounded the 
proof of the four color theorem in the seventies, which involved the use 
of computers doing parts of the proof that would be beyond 
human capability \cite{w02}. 
This time around, however, the questions raised need answers.  
Be as it may, some large pinch of caution is needed.

%----------------------------------------------------
\section{The power of Insight}
\label{insight} 

One of the main criticism to most intensive ML applications, such as AlfaFold and
even more so latest chatGPT Large Language Model (LLM) algorithms, is the astronomical 
number of weights used, rapidly moving into the hundred billion regime.
This is problematic in many respects, both fundamental and practical.

The fundamental aspect is that in theoretical physics,  parameters 
are traditionally held as fudge factors, i.e., temporary fixes for our
holes of understanding. Hence, the fewer parameters the better. 
Newton's law of gravitation epitomizes the beauty and 
universality of a good theory. Once you understand that two 
material bodies attract with a force proportional to the product of 
their masses and inversely proportional to the square 
of their distance, all you need to fix  by experiment is the ratio 
\begin{equation}
G = \frac{F r^2}{m_1 m_2}
\end{equation}  
The beauty of this expression rests with its universality: {\it any\/} experiment, whether you
are using apples, billiard balls or the moon orbiting around planet Earth, 
will return the {\it same\/} value for this ratio, 
$G \approx 6.67 \times 10^{-11}\ Nm^2/kg^2$. 
All data, big or small, are captured within a single parameter (as long as
gravitation is weak enough and non-quantum)!
This is the power of Insight and explains why physicists 
place so much value in it.
To this regard, it is worth recalling the famous Fermi's anecdote
reported by Freeman Dyson, when he approach Fermi to
discuss with him his pseudo-scalar theory for pions.
Fermi asks ``how many parameters do you have in your model?" "Five" replies
Dyson. ``My friend John von Neumann told me that with four parameters he can fit 
an elephant, and with five he can wriggle his trunk."
And with that the conversation was over \cite{dyson}.

The practical aspect has to do with sustainability: it is estimated that training
next-generation chat-bots with some hundreds billion parameter may easily move into
the GWatt power demand, corresponding to the output from a substantial nuclear power plant.
The comparison with the ten Watts of our brain is embarrassing, but that is
another story \cite{LLMScaling}. 

One could observe that present-day top-end supercomputers are 
also pretty energy-thirsty with a power request in the 
order of ten MWatts, a million times more than human brain.
The point though is that there a return for this: exascale 
computers compute some twenty orders of magnitude faster than 
our brain (a tiny fraction of Flops/s). 
So, the question becomes social and ethical: is it worth 
exhausting a substantial fraction of the worldwide 
energy budget to feed the insatiable appetite of chat-bots?

Science-wise, the astronomical disparity between the LLM's power request
and that of our brain provide a strong pointer towards 
the need for a much better theory of machine learning \cite{h22}.
This may spawn a genuinely new way of doing science but
to achieve this goal it is important to keep an open mind.
Here comes the point.

As noted above, centuries of physics (since Galileo)
have taught us that the least number of parameters the best.
So, let us call $P$ the number of parameters required to produce
a satisfactory fit to an ensemble consisting of $D$ data.
We may define a fitting efficiency as the ratio
\begin{equation}
f = \frac{D}{P}\;.
\end{equation}
Clearly, the scientific method aims at large 
values of $f$, the zero-parameter limit $f \to \infty$ denoting 
the "Perfect Theory", one with no free-parameters at all.
The Standard Model, still our most accurate description of fundamental interactions,
is regarded by some as "ugly" because it demands 19 free parameters. 

Machine learning, and most notably LLM's, work instead in the opposite
limit $f \ll 1$, $f<1$ denoting the infamous ``over-fitting" regime: more
parameters than data. 
Over-fitting is a notorious problem for ML, as it hinders the capability
of extrapolate to capture unseen data, the whole purpose of the ML ordeal.
Fact remains, though, that ML practitioners have proven capable of
turning around it under circumstances where it was supposed
to hit hard, the famous google transformer paper being 
an adamant example in point \cite{v17}:
No systematic theory, a mathematical framework based on 
billions of free parameters, augmented with a set of 
semi-empirical hunches and recipes. 
Yet, in the end, it often works and sometimes big time so.
Hence, while it is entirely healthy to remain skeptical of 
occasional success stories, no matter how spectacular, one should 
also be open to the possibility of making the most of
this "unsuspected" capability  of machine learning
to bypass overfitting.

%----------------------------------------------------
\section{Fermi's belt in Los Alamos}
\label{belt}

This said, the Nobel Prize for Chemistry raises a point which goes
beyond science. 
Let us assume that from now on that we have passed the Singularity
and live in a world where radical Empiricism takes the lead
in science: Correlation does supersede Causation and science
can advance with just a little ancillary help of theory and simulation.
Let us say that this is the ugly but effective winning route.

Our point here that at a deeper level, this  is no 
longer about Correlation versus Causation, but 
rather Control versus Insight.

Current chat-bots can write pretty decent code in seconds, in the face
of the hours or days for a skilled programmer.
You could hail at this as to a major time-saver and for sure it is.
But if you dig just a bit deeper, a poisoned apple pops out in plain sight.

Let's go back to Fermi again.
There is little question that Fermi was one of the most versatile
physicist of all time, with several achievements under his belt each worth
a Nobel prize (fission, Fermi-Dirac statistics, beta decay, Thomas-Fermi theory of nuclei etc.). 

Amazingly, once interviewed about what he regarded as his 
most impressive achievement, none of these were to come up in his reply.
Instead, he quoted an episode from the Los Alamos period, when his Jeep got 
stuck in the desert because the transmission belt went bust.
Not your  best cup of tea if you are left alone in the Los Alamos desert...
Fermi being Fermi, he managed to get out of the hook by replacing
the transmission belt with ... his own belt!

That means being able to face tough problems and adversity, something
that the relentless promise to relieve us, in fact our brain,
from any burden, is rapidly grinding to a halt.
A few high-tech companies will take care of writing codes for you, that
is where Control of a few over the rest of us, shows up beyond any
reasonable doubt: chat-bots write codes in seconds, no need of 
programming for the new generations, Google brains will take care of this for you.

Now, leaving aside Fermi, even on our modest personal scale, we can quote
many instances in which we were sure we had it all sorted out, but when
it was time to finally code it up, we actually realized that we did not 
really know how to do it exactly. 
So we had to pedal our way back and figure out where the loophole was, a 
process of immense value for our scientific growth.

One may see a concrete example of this in comparing 
References \cite{h16} and \cite{h18}. 
The first paper, based entirely on theory, missed a crucial ingredient 
--- the co-moving velocity --- which is of increasing importance in 
the field of two-phase flow in porous media. 
It was only through computations that we discovered it, and with 
it in hand, we realized that Reference \cite{h16} was not correct, replacing it with \cite{h18}.

So, the question is: is it worth saving our time to code it
up? Our answer is yes, but only to a point; We surely welcome 
automatic help, but not to the point where it would kick us
completely out of the show.

We all speculate about the Singularity as the day when AI supersedes Natural Intelligence,
(NI) where the latter is typically thought of a constant in time.
It is not, at least on average, some of the most aggressive AI 
applications do lower NI levels, thereby accelerating the Singularity. 
Hence, the Singularity itself is probably not as much of a problem as 
the degradation of NI. This is not to say that AI is should be rejected, but
rather to stand for a cooperative pattern whereby the ultimate
control is left to NI. 

%----------------------------------------------------
\section{The bright side of machine learning}
\label{grumpy}

In the previous sections we have raised a number of warnings against the 
backsides of AI, and particularly to its most aggressive claims.
The ``grumpy old men" part of the paper ends here.

Indeed, it would be poor-sighted to deny that machine-learning 
has brought a new dimension to scientific investigation 
along the three standard pillars of Theory, Simulation
and Experiment. AI is contributing in many respects to the scientific endeavor but here
we focus on one that appears to be  particularly relevant 
across many scientific and societal applications:
the infamous Curse of Dimensionality (CoD) \cite{COD}.

Our brain, as well as much of our math, is notoriously 
at odds with handling high-dimensional information,
the main reason being the exponential growth of the volume with dimensions
$V_D = V_1^D$. Assuming a uniform density of information, $\rho$,
the total information $I(V)=\rho V$ stored in a given 
volume of state space also scales exponentially with the number of dimensions. 
Take standard four-dimensional spacetime, with $N=10^3$ degrees
of freedom per dimension and unit density everywhere, we obtain 
$I_4=10^{12}$, which is basically as much as we can accommodate 
on present-day Exascale computers.
Many problems in science and engineering live in much higher dimensional
state spaces with thousands, millions or even billions of dimensions, spelling
complete doom for most of our mathematical methods.
Fortunately, Nature is usually merciful and the amount of information does 
not grow accordingly because most of these ultra-dimensional spaces are
empty and the relevant (active) degrees of freedom use to populate a 
much smaller  manifold of dimension $d<<D$ ($d$ is usually known as Intrinsic
Dimension). Finding such manifolds is a highly non-trivial task, not 
only because they occupy an extremely tiny portion of state-space
but also because their topology is often highly irregular and scattered out. 
This is a central issue in modern computational statistical physics 
and many other fields of modern science.      
Machine learning in general and transformers in particular can be viewed
as highly non-ergodic discrete dynamical systems, targeted to locate
the solution manifold as efficiently as possible, without wasting resources
to visit empty regions of state-space.  
Let us spell the idea out in some more detail.
Transformers operation can be paralleled to a discrete
dynamical system of the form \cite{NPDE}:
\begin{equation}
\label{FW}
y = f^L[Wx-b]\;,
\end{equation}
where $x$ is the input state in $D$-dimensional feature space, $y$ is the
corresponding output, the pair $(W,b)$ indicates the set of 
weights and biases connecting the hidden layers $\lbrace z_1, z_2 \dots z_L \rbrace$
and $f$ is the activation function applied across each of the $L$ hidden layers.
The above "forward-step" is complemented by a backward-error propagation step
in which the weights are adjusted in such a way as to minimize the departure
from the desired target ("truth") $y_T$, also known as Loss Function
\begin{equation}
\label{BW}
\mathcal{L} = ||y-y_T||\;,
\end{equation}
where $||.||$ indicates a suitable metric in feature space.
Such minimization is usually performed with steepest-descent-like techniques
\begin{equation}
\label{BW2}
W' = W - \alpha \frac{\partial \mathcal{L}}{\partial W}\;,
\end{equation}
where $\alpha$ is a relaxation parameter known as "learning rate".
The above backward-forward loop is repeated over a huge set of data
$(x,y)$ until the optimal weight configuration is found.

% -------------------------
\begin{figure}
\centering
\includegraphics[scale=0.2]{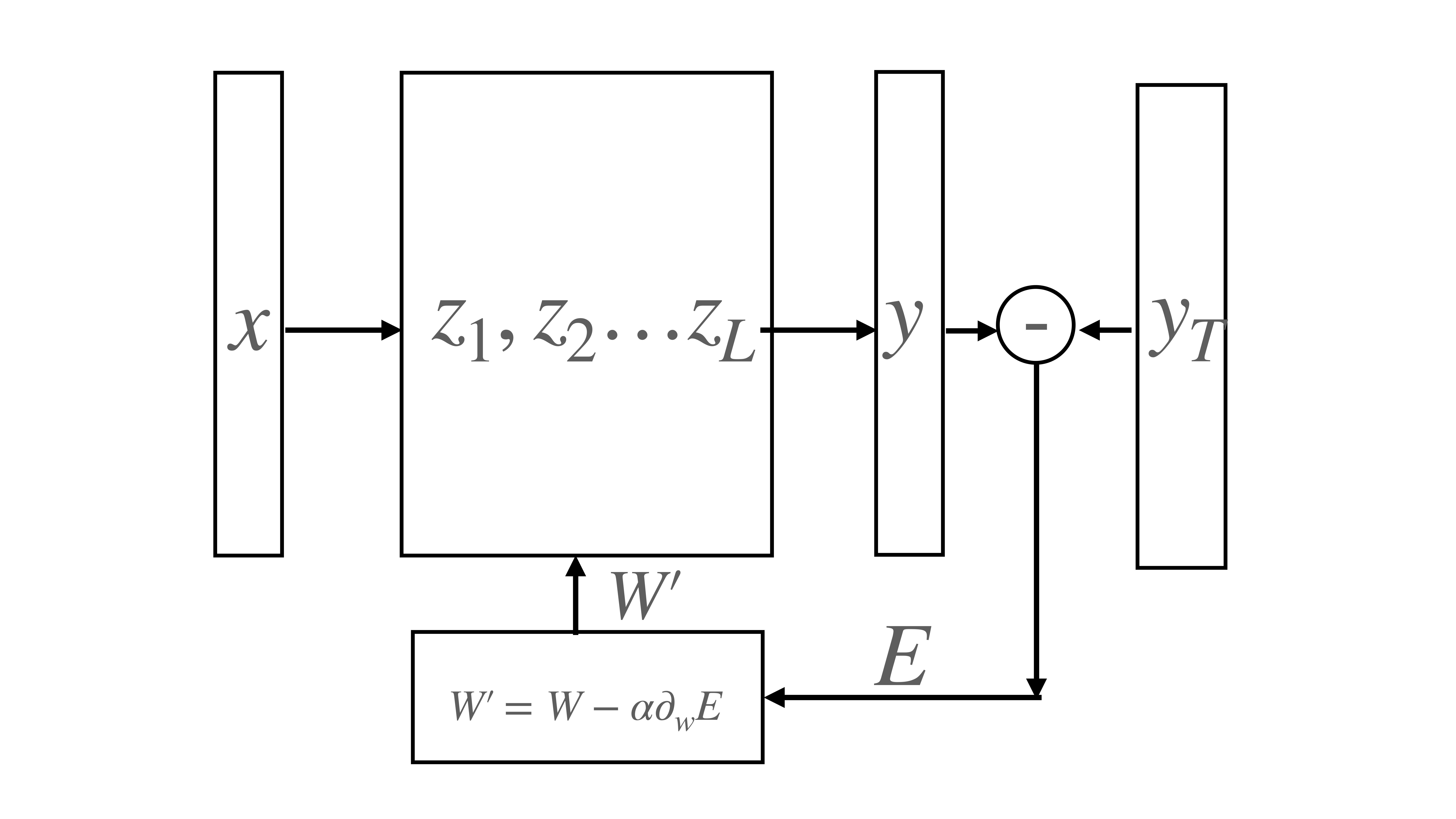}
\caption{
A schematic representation of the transformer architecture.
Note that each layer carries the same number of neurons.
}
\end{figure}
% -------------------------

By paralleling the layers to discrete time steps, the 
above procedure amounts to a discrete dynamical system starting 
at $z_0=x$ and ending at $z_{L+1}=y$, evolving under the
feed-back control of the backward error propagation step. 

The analogy has been discussed in detail in \cite{NPDE} and here
we only point out that such trajectories appear to be 
efficient in catching the desired target $y_T$ in ultra-dimensional space.
For instance leading-edge LLM's with near trillions weights can find solutions
in manifolds with $d \sim 40$.
Leaving aside all reservations about computational and energetic 
parsimony as well as lack of physical insight and explainability, this is 
unquestionably a remarkable deed.

The magic is probably less arcane than it may seem at first sight.
The transformer loop discussed above typically lands on random matrix
solutions for the weights, even in the case where the problem has
a definite structure, say a sparse matrix using standard discretization 
techniques. This may seem weird at first glance, but it actually
reflects the fact that the ensemble of random matrices is astronomically
larger than the ensemble of ordered (structured) matrices arising
from grid discretization methods.
It is therefore no surprise that machine-learning search ends up in this
huge ensemble rather than on the incommensurably  smaller ensemble 
of structured matrices. In the end, the usual entropic argument. 
The price, of course, is a vastly larger number of parameters and training
costs, aggravated by the lack of a systematic convergence theory.
Ignoring the latter dark-sides, as it is typical of the most aggressive 
AI claims, is to be highly deprecated and must be countered: the 
Fermi's belt anecdocte should not go forgotten.
Yet, the fact remains that developing suitable strategies combining 
the conceptual transparency
of the scientific method with the ability of transformers to chase "golden
nuggets" in ultra-dimensional spaces, holds major potential to transformative
progress in the way scientific investigation will be pursued in the years to come.   

%----------------------------------------------------
\section{Outlook}
\label{outlook}

Machine Learning is often hyped as a universal panacea, which
is most certainly not. Hence it is crucial to keep a critical 
attitude towards the most bombastic and aggressive AI claims. 
However a judicious and clever combination of computer
simulation based on physical insight and the machine learning power
to explore active regions of ultra-dimensional spaces, may lead to 
transformative progress in the future of science.

%----------------------------------------------------
\bigskip
\small 

This work was partly supported by the Research Council of Norway 
through its Centers of Excellence funding scheme, project number 262644. 
AH acknowledges funding from the European Research Council (Grant Agreement 101141323 AGIPORE).
SS acknowledges funding from the from the Research and Innovation programme "European Union’s
Horizon Europe EIC pathfinder" under grant agreement No101187428”.
%Views and opinions expressed are however those of the
%author(s) only and do not necessarily reflect those of
%the European Union or European Innovation Council
%and SMEs Executive Agency (EISMEA). Neither the
%European Union nor the granting authority can be held
%responsible for them."
%----------------------------------------------------

%----------------------------------------------------
\end{document}